\documentclass[prl,twocolumn,preprintnumbers,amsmath,amssymb,superscriptaddress]{revtex4}

\usepackage{graphicx}
\usepackage{dcolumn}
\usepackage{bm}
\usepackage[latin1]{inputenc}
\usepackage{color}

\newcommand{\mean}[1]{\langle #1\rangle}

\begin{document}

\title{Dynamics of formation and decay of coherence in a polariton
  condensate}

\author{E. del Valle}

\affiliation{School of Physics and Astronomy, University of
  Southampton, SO171BJ, Southampton, UK.}

\author{D. Sanvitto} 

\affiliation{Dep. F\'{\i}sica de Materiales, Universidad Aut\'onoma de
  Madrid, 28049, Madrid, Spain.}

\author{A. Amo} 

\affiliation{Laboratoire Kastler Brossel, Université Pierre et Marie Curie, Ecole Normale Supérieure et CNRS, UPMC Case 74, 75252 Paris Cedex 05, France.}

\author{F.P.~Laussy} 

\affiliation{School of Physics and Astronomy, University of
  Southampton, SO171BJ, Southampton, UK.}

\author{R. Andr\'e} 

\affiliation{CEA-CNRS. Institut NEEL-CNRS, BP166. 38042 Grenoble Cedex
  9, France.}

\author{C. Tejedor}

\affiliation{Dep. Física Teórica de la Materia Condensada, Universidad
  Autónoma de Madrid, 28049, Madrid, Spain}

\author{L. Vi\~na} 

\affiliation{Dep. F\'{\i}sica de Materiales, Universidad Aut\'onoma de
  Madrid, 28049, Madrid, Spain.}

\date{\today}

\begin{abstract}
  We study the dynamics of formation and decay of a condensate of
  microcavity polaritons. We investigate the relationship between the
  number of particles, the emission linewidth and its degree of
  linear polarization which serves as the order parameter. Tracking
  the condensate formation, we show that coherence is not determined
  only by occupation of the ground state, bringing new insights into
  the determining factors for Bose-Einstein condensation.
\end{abstract}

\maketitle

Whereas a lot is known of the thermodynamics of quantum fluids---at
equilibrium and in infinite size systems---the dynamics of formation
of a Bose-Einstein condensate (BEC) is still vastly an open
question. This dynamics is not easily detectable in atomic BEC and has
not been studied experimentally, due to the very short timescale in
which cold atomic gases reach thermal equilibrium. A good alternative
to investigate the process of condensate formation is that of
microcavity-polaritons~\cite{weisbuch92a}, although this system
differs from atomic condensates due to its intrinsically
out-of-equilibrium character.  Polaritons arise from the strong
coupling of photons and electrons in semiconductor
microcavities~\cite{kavokin_book07a} and have recently shown to
undergo a non-equilibrium phase transition with a spontaneous buildup
of coherence in the ground
state~\cite{kasprzak06a,balili07a,christopoulos07a,baumberg08a,bajoni08a},
similar to what has been observed in atomic
BEC~\cite{anderson95a,davis95a}.  Thanks to the small polariton
lifetime, various properties can be tracked continuously from the
photons they emit. This has been used to measure the spatial first
order coherence~\cite{deng07a}, the temporal first and second order
correlation functions~\cite{kasprzak08a,love08a} and even real time
dynamics such as propagation~\cite{amo09a}.

The dynamical evolution and origin of the degree of coherence in a
polariton condensate remain however unclear.  Intriguing results have
been recently reported of one order of magnitude increase of the
ground state polaritons coherence time, by reducing particle
fluctuations in the excited states~\cite{love08a}. Such experiments
reveal the possibilities for the dynamical observation of
condensation~\cite{nardin}. Also some theoretical studies have been
reported based on stochastic simulations of the order
parameter~\cite{fsread08a}.  In this work, we pursue this goal both
experimentally and theoretically, and track the dynamics of formation
of a polariton BEC, following how the condensation forms and decays in
a pulsed experiment.

Since polaritons have a spin degree of freedom that is passed to the
polarization of the photon they emit, the buildup of off-diagonal
elements in the density matrix (characteristic of coherent states) can
be accessed through the degree of linear polarization:
\begin{align}
  \label{eq:degree}
  D_\mathrm{l}={2\mid
    \Re{\mean{a_{0\uparrow}a_{0\downarrow}^\dagger}}\mid}/
  \big({\mean{a_{0\uparrow}^\dagger
      a_{0\uparrow}}+\mean{a_{0\downarrow}^\dagger
      a_{0\downarrow}}}\big)\,,
\end{align}
with $a_{0 \uparrow}$ and $a_{0 \downarrow}$ the field operators for
spin-up (right-circular polarization) and spin-down (left-circular
polarization) polaritons, respectively. If spin-up and spin-down
polaritons are uncorrelated, with a thermal distribution and no
determined phase relationship, the light is unpolarized. If, however,
the two spin-projections are condensed with well defined order
parameters~$\langle a_{0\uparrow}\rangle$, $\langle
a_{0\downarrow}\rangle$ that factorize the numerator of
Eq.~(\ref{eq:degree}), the emitted light acquires a definite
polarization. Therefore, $D_\mathrm{l}$ is linked to the degree of
coherence of the condensate and to the imbalance of spin-up/spin-down
polaritons~\cite{laussy06a,shelykh06a,kasprzak08a,baumberg08a}. The
spin degeneracy can be lifted producing a Zeeman splitting of
polariton energies. If we can further assume that the two spin
populations have similar evolutions, these can be modelled with a
convolution of thermal and coherent states~\cite{lachs65a,laussy06a}
with time-dependent thermal and coherent fractions. In our case, we
have observed experimentally that, except for a short transient after
the pulsed excitation, spin-up and spin-down populations equalize and
evolve similarly in time~\cite{Martin02a}. Therefore the degree of
linear polarization is directly related to the degree of second order
coherence (at zero delay)~$g^{(2)}$ of the polariton gas:
$D_\mathrm{l}\approx\sqrt{2-g^{(2)}}$. If the spin populations have
very different dynamics, one has to keep track of the coherence degree
of each of the two spin components and this simple expression no
longer holds. Up to now, this relationship has been exploited for
measuring the degree of coherence only in stationary
situations~\cite{kasprzak08a,baumberg08a}. In this work, we use it to
study experimentally the dynamics of condensation. Using simultaneous
detection of energy- and time-resolved photoluminescence, together
with the degree of linear polarization of the ground state, we monitor
the formation and decay of the polariton condensed phase in the ground
state after a circularly-polarized, pulsed excitation at high
energy. Comparing our data with a phenomenological model, we
investigate the evolution of the condensate statistics and coherence.

\begin{figure}[t]
\includegraphics[width=\linewidth]{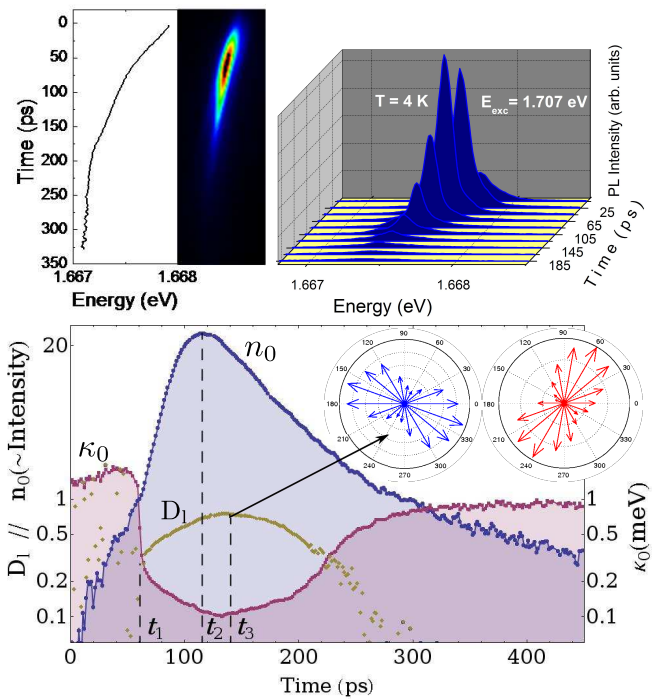}
\caption{(Color online) Experimental results on the dynamics of BEC
  formation after the arrival of the non-resonant pulse. (a) Emission
  energy from the ground state ($k=0$) as a function of time (left)
  and the corresponding photoluminescence intensity (right). The color
  code goes from black, no emission, to red, intense emission,
  saturating in a black central region. (b) Spectral lineshapes
  extracted from~(a).  (c) In blue dots, average condensate
  population~$n_0$ (normalized photoluminescence intensity); in purple
  squares, linewidth~$\kappa_0$ (meV) of the emission peak, limited by
  the spectral resolution of~$\sim0.1$~meV; in brown rhombus, degree
  of linear polarization $D_\mathrm{l}$. Times for coherence build-up
  ($t_1$), maximum population ($t_2$) and coherence ($t_3$) are marked
  with dashed lines. Inset: polarization polar plot at $t=t_3$ with
  maximum coherence for the present case (I) and for a second point in
  the sample~(II).}
\label{fig:1}
\end{figure}

In the experiments, polaritons are created in a CdTe-based microcavity
grown by molecular beam epitaxy. The sample consists of a
Cd$_{0.4}$Mg$_{0.6}$Te $2\lambda$-cavity with top (bottom) distributed
Bragg reflectors of 17.5 (23) pairs of alternating $\lambda/4$-thick
layers of Cd$_{0.4}$Mg$_{0.6}$Te and Cd$_{0.75}$Mn$_{0.25}$Te. Four
CdTe QWs of 50~\AA~thickness, separated by a 60~\AA~barrier of
Cd$_{0.4}$Mg$_{0.6}$Te, are placed in each of the antinodes of the
electromagnetic field. Exciton-photon coupling is achieved with a Rabi
splitting of $23$~meV; the sample is kept at $4$~K and condensation is
obtained around zero detuning between the exciton and the bare cavity
dispersion. Photons are injected via circularly polarized, ps-long
pulses tuned at the first minimum above the Bragg mirror stopband
($\sim40$~meV from the bottom of the dispersion) in order to guarantee
the memory loss of the laser polarization and the coherence by the
time polaritons are formed at the bottom of the band. Light escaping
from the microcavity is dispersed through a~$0.5$ m spectrometer and
analyzed with a streak camera. This allows to resolve in time the
intensity and energy of the emission of the ground state
(Fig.\ref{fig:1}(a) and (b)), and simultaneously analyze the direction
and linear degree of polarization of the emitted light together with
the polariton linewidth (Fig.\ref{fig:1}(c) and insets). The direction
of the linear polarization is pinned to a crystallographic axis, as
has been observed with continuous pumping in a sample of the same
batch growth~\cite{kasprzak08a} and with pulsed pumping in samples
similar to ours~\cite{Klopotowski06a}. Without this pinning, the
polarization would average to zero, which is detrimental to our
approach but is otherwise one of the most direct demonstration of
symmetry breaking associated to the quantum phase transition, as
reported by Baumberg \emph{et al.}~\cite{baumberg08a}.  In our case,
the direction of the pinning depends on the point of the sample, as
shown in the insets~(I) and~(II) of Fig.\ref{fig:1}(c).

\begin{figure}[tb]
\includegraphics[width=.8\linewidth]{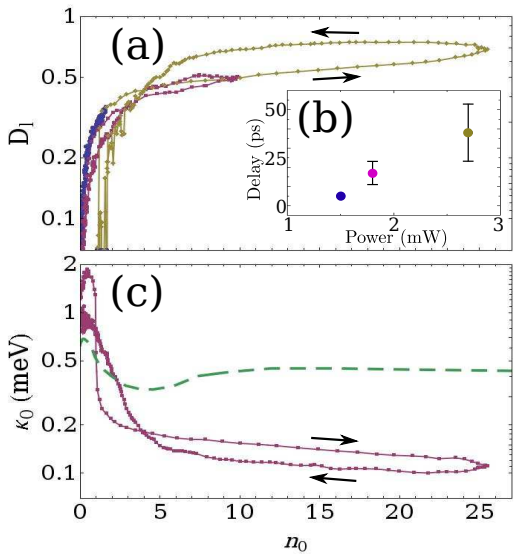}
\caption{(Color online) Hysteresis in $D_\mathrm{l}$~(a) and
  $\kappa_0$~(c) as a function of $n_0$, with time running in the
  sense of the arrows.  (a) Three experiments are shown with
  increasing excitation powers (the one under study in brown
  rhombus). (b) Growth with power of the delay between maximum
  population and coherence for these three cases.  The
  linewidth of Ref.~\cite{kasprzak06a}, under CW excitation, is
  given in (c) (dashed green) for comparison.}
\label{fig:2}
\end{figure}

The non-resonant laser excitation creates a large population of
carriers, at high energies, that form the polaritons.  Thanks to the
quick relaxation mechanisms, like phonon emission and
polariton-polariton scattering~\cite{doan05a}, the population of
polaritons relaxes into the fundamental state (with in-plane momentum
$k=0$). Figure~\ref{fig:1} depicts the experimental time evolution of
the ground state after the arrival of the pulse at some time $t<0$,
till the complete decay of the condensate. The intensity of the light
emitted by the condensate, first shown in Fig.~\ref{fig:1}(a)-right
(in density plot), is normalized to represent the ground state
population $n_0$ in Fig.~\ref{fig:1}(c) (blue circles). The abrupt
change into a nonlinear growth---associated with the onset of
stimulated enhancement---is matched to~$1$.  We note from the ground
state energy (Fig.~\ref{fig:1}(a)), that the dispersion is already
blueshifted (as compared to the long time energy, at $t\geq 300$~ps),
even when $n_0$ is very small (at $t\simeq 0$). This confirms that the
blueshift is mainly due to exciton screening and carrier-carrier
interaction in the exciton
reservoir~\cite{peyghambarian84a,schmittrink86a,ciuti00a}, and that it
is only slightly affected by the ground state occupation. Note that
the blueshift is always much smaller ($<1$~meV) than the Rabi
splitting, so that the system remains in strong-coupling. As the
reservoir depletes with time, the overall population and the blueshift
monotonically decrease, while $n_0$ first increases and then
decreases.  The coherence imprinted by the laser pulse and its
original polarization is lost during this efficient cascade of decays,
populating the ground state with initially uncorrelated
polaritons. This is clear from the linear polarization $D_\mathrm{l}$
(brown rhombus in Fig.~\ref{fig:1}(c)), that takes some time to build
up after $n_0\approx 1$. Passed this particular time (corresponding to
$t_1=61$~ps in the figures) the system is above threshold with a
condensed fraction and a phase locking of the spin-up and spin-down
condensates. Stimulated scattering into the ground state becomes the
dominating process and the population $n_0$ grows up to a maximum
(with $\max(n_0)\approx25$ at~$t_2=118$~ps). However, the coherence
reaches its maximum, $\max(D_\mathrm{l})=0.75$, at a later time,
$t_3=140$~ps, with a proper dynamics that does not follow
instantaneously that of $n_0$: $D_\mathrm{l}$ decays for another
$150$~ps at a different rate than that of the population.
$D_\mathrm{l}$ is represented as a function of $n_0$ in
Fig.~\ref{fig:2}(a), and compared with two other cases in order to
make evident the existence of an hysteresis loop. As the excitation
power is decreased, not only $n_0$ and $D_\mathrm{l}$ decrease, but
the hysteresis also gradually disappears.

The linewidth of the ground state emission is also dramatically
affected by the condensation. In the simplest picture of a single mode
with an effective pumping rate from all the other
states~$W^\mathrm{out}_0=\sum_k P_{k0}(1+n_k)$ ($n_k$ the population
of the $k$th state) and a radiative decay rate~$\gamma_0$ at $k=0$,
the linewidth can be approximated by
$\kappa_0=(W^\mathrm{out}_0+\gamma_0)/(1+n_0)$. This expression is
enough to understand qualitatively the linewidth dynamics. As shown in
Fig.~\ref{fig:1}(c) (purple squares), $\kappa_0$ drops dramatically at
the phase transition due to the exponential growth of $n_0$. Close to
the maximum population, the linewidth reaches the limiting value of
our spectral resolution ($0.1$~meV). Note that $\kappa_0$ not only
depends on $n_0$ but is also affected (broadened) by the populations
of the levels feeding the ground state, which are enhanced before the
transition takes place.  The steady state value of
$\kappa_0\rightarrow 0.9$~meV ($t>300$~ps) corresponds to the
linewidth of the bare ground state polaritons, $\gamma_0$. This
results in another hysteresis shown in Fig.~\ref{fig:2}(c).  Although
polaritons, being interacting bosons, also suffer screening due to
Coulomb interaction, this effect is only significant at extremely high
polariton densities~\cite{porras03a}. This regime is never reached in
our experiment.  Kasprzak \emph{et al.}~\cite{kasprzak06a} observed
under CW excitation an increase in the linewidth (dashed, green line
in Fig.~\ref{fig:2}(c)) and attributed it to polariton-polariton
interaction. However, we find that the ground state population is too
small to produce this effect (the particle number should be three
orders of magnitude higher). The reason for such a significant
increase of the linewidth is more likely to be found in the high
number of reservoir particles, much higher than $n_0$ in the case of
CW experiments. With our time dependent excitation, excluding the
first few tenths of ps, the amount of particles in the reservoir is
significantly smaller than that at the bottom of the polariton
dispersion, reducing drastically this extra decoherence effect.

\begin{figure}[t]
  \includegraphics[width=\linewidth]{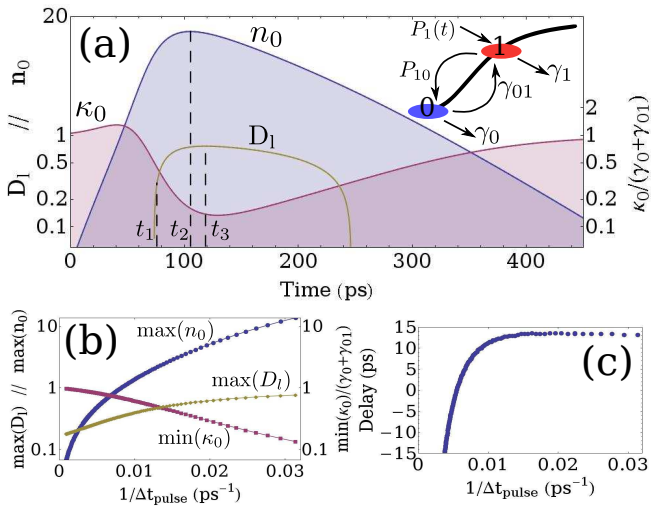}
  \caption{(Color online) (a) Results from the theoretical model
    (sketched in inset): $n_0$ (blue), $D_\mathrm{l}$ (brown) and
    $\kappa_0$ (purple), as a function of time (rescaled to ps for
    comparison).  In (b), the maximum $\max(n_0)$,
    $\max(D_\mathrm{l})$ and minimum
    $\min(\kappa_0)$ values reached in the simulations as the
    effective pulse narrows (with increasing~$1/\Delta
    t_\mathrm{pulse}$) down to the case in (a). The delay $t_3-t_2$
    between $\max(n_0)$ and $\max(D_\mathrm{l})$ is
    also plotted in (c).}
  \label{fig:3}
\end{figure}

Our experiments on polariton coherence formation are supported by a
simple two-level model (sketched in the inset of Fig.~\ref{fig:3}(a)),
that links together the dynamics of the coherence degree with the
population and the linear polarization. We work under the
assumption---supported by the observation---that both spin components
evolve similarly following this dynamics.  We proceed in the spirit of
the evaporative cooling description~\cite{holland96a}, that has been
used before in models including polariton-polariton and
photon-polariton scattering~\cite{porras03a,laussy04c}. A reservoir,
state~$1$, is populated in an incoherent way from higher levels, what
we describe with an effective time-dependent rate~$P_1(t)$ which has a
Gaussian profile in time, reminiscent of the excitation of the laser
at very high energy, with temporal effective width~$\Delta
t_\mathrm{pulse}$. These particles are scattered incoherently into the
ground state, level~0, at a different rate $P_{10}$. Both levels loose
polaritons with decay rates $\gamma_{0}$ and $\gamma_1$, respectively.
We also include the inverse process of promoting polaritons up the
branch at some small rate $\gamma_{01}$.  This dynamics is described
with Lindblad terms in a master equation of the ground-state/reservoir
system. This leads to rate equations for the time dependent
distributions of particles, which are solved numerically to compute
the observables of interest ($n_0$, $g^{(2)}$, $D_\mathrm{l}$,
etc\dots).  The linewidth is obtained from
$\kappa_0\approx[\gamma_{01}(1+n_1)+\gamma_0]/(1+n_0)$.  In this
model, the coherence builds~up spontaneously, while other
works~\cite{rubo03a} require a seed (such as an initial coherent
state) in order to investigate the dynamics of coherence. The results,
plotted in Fig.~\ref{fig:3}(a) with the same color code, are
qualitatively similar to the experimental data in
Fig.~\ref{fig:1}(c). With no need for more levels, we also reproduce
the existence of a delay between maximum coherence and population. We
find that when the effective pulse $P_1(t)$ gets sharper in time
(small $1/\Delta t_\mathrm{pulse}$), not only $\max(n_0)$ and
$\max(D_\mathrm{l})$ are larger, but also the delay between them
increases, as can be seen in Figs.~\ref{fig:3}(b) and (c). The delay
can even become negative when the pulse is too flat. The origin and
sign of the delay can be explained as follows: for a sharp and strong
pulse, the ground state population $n_0$ rises so quickly that
coherence formation cannot simultaneously build~up (through stimulated
interaction with the upper level), and settles to its own, intrinsic
dynamics instead (the delay becomes independent of the pulse duration
as this one gets narrower).  With a flat pulse, on the other hand,
$n_0$ grows slowly and weakly, giving time to the condensate coherence
to form together with it, possibly overtaking the dynamics for very
long effective pulses and yielding a negative delay
(Fig.~\ref{fig:3}(c)).  Increasing the in-flow of particles (e.g.,
increasing $P_{10}/\gamma_{01}$), also has a similar effect of
enhancing the coherence and its rising time. On the other hand, the
total intensity of the pulse is not determining the delay.  This
argument can be extended to the experiment, where a more intense pulse
also results in a sharper profile for the effective excitation, given
that the chain of de-excitations towards level~1 is more efficient,
thanks to the density dependence of the exciton-exciton
scattering~\cite{munoz95a}.

Finally, although our simple model does not lead to a quantitative
agreement with the experiments (time has been rescaled for a better
comparison), it is of a fundamental character, since it contains the
minimum number of elements to reproduce BEC formation and links
together its essential features (population, coherence, linewidth and
polarization). The model should include more levels in the
de-excitation chain towards the ground state to reach to a
quantitative agreement, not bringing, however, any essential new
features into the description (as we have checked but which we do not
discuss here).

In conclusion, we report the experimental observation of the dynamics
of a polariton condensate, created under a non-resonant, circularly
polarized pulse, and the study of its coherence buildup and decay. We
track the order parameter of the transition through the degree of
linear polarization, that spontaneously builds up. This occurs when
the average population of the ground state exceeds one and is
accompanied by an abrupt decrease in the emission linewidth.  The
maximum degree of coherence that is achieved does not coincide in time
with the maximum population, evidencing that the coherence of the
condensate has its own particular dynamics.  The pulsed excitation
regime is a worthy tool revealing neatly the interplay of all the
relevant quantities of polariton condensates and displaying the
mechanisms of formation and decay of a BEC. We support our claims with
a simple theoretical model.

\begin{acknowledgments}
  We thank D.~Porras for fruitful discussions and the Spanish MEC
  (MAT2008-01555/NAN, QOIT-CSD2006-00019), CAM (S-0505/ESP-0200) and
  the IMDEA-Nanociencia for funding. EdV acknowledges the Newton
  Fellowship program and DS the Ram\'on y Cajal program.
\end{acknowledgments}

\bibliography{Sci,books,Elena}

\end{document}